\documentclass[conference]{IEEEtran}

\usepackage[english]{babel}

\RequirePackage[l2tabu, orthodox]{nag}

\usepackage{pdflscape}

\usepackage{newtxtext,newtxmath}

\usepackage{cite}

\usepackage{booktabs}
\newcommand{\ra}[1]{\renewcommand{\arraystretch}{#1}}

\usepackage{multirow}

\usepackage{caption}

\usepackage{tabularx}
\usepackage{xcolor}

\definecolor{myblue}{HTML}{0C5DA5} 
\definecolor{mygreen}{HTML}{00B945} 
\definecolor{myorange}{HTML}{FF9500} 
\definecolor{myred}{HTML}{FF2C00} 
\definecolor{mypurple}{HTML}{845B97} 
\definecolor{mygray}{HTML}{474747} 

\usepackage[pdftex]{graphicx}
\graphicspath{{./figures/}{./resources/}}
\DeclareGraphicsExtensions{.pdf,.png,.jpg,.jpeg}

\usepackage{tikz}
\usetikzlibrary{shapes,arrows}
\usepackage{pgfplots}
\pgfplotsset{compat=1.9}

\usepackage{amsmath}
\usepackage{mathtools}
\DeclarePairedDelimiter{\norm}{\lVert}{\rVert}

\ifCLASSOPTIONcompsoc
  \usepackage[caption=false,font=normalsize,labelfont=sf,textfont=sf]{subfig}
\else
  \usepackage[caption=false,font=footnotesize]{subfig}
\fi

\usepackage{url}

\usepackage{xspace}
\makeatletter
\DeclareRobustCommand\onedot{\futurelet\@let@token\@onedot}
\def\@onedot{\ifx\@let@token.\else.\null\fi\xspace}

\def\eg{\emph{e.g}\onedot}

\def\ie{\emph{i.e}\onedot}

\def\etal{\emph{et al}\onedot}
\makeatother

\begin{document}

\title{Improving CSI-based Massive MIMO Indoor Positioning using Convolutional Neural Network}

\author{\IEEEauthorblockN{%
    Gregor~Cerar\IEEEauthorrefmark{1}\IEEEauthorrefmark{2},
    Ale\v{s}~\v{S}vigelj\IEEEauthorrefmark{1}\IEEEauthorrefmark{2},
    Mihael~Mohor\v{c}i\v{c}\IEEEauthorrefmark{1}\IEEEauthorrefmark{2},
    Carolina~Fortuna\IEEEauthorrefmark{1},
    and 
    Toma\v{z}~Javornik\IEEEauthorrefmark{1}\IEEEauthorrefmark{2}
}\IEEEauthorblockA{%
    \IEEEauthorrefmark{1}Department of Communication Systems, Jo\v{z}ef Stefan Institute, SI-1000, Slovenia.\\
    \IEEEauthorrefmark{2}Jo\v{z}ef Stefan International Postgraduate School, SI-1000, Slovenia.
}%
\{gregor.cerar $\mid$ ales.svigelj $\mid$ miha.mohorcic $\mid$ carolina.fortuna $\mid$ tomaz.javornik\}@ijs.si
}


%



\maketitle

\begin{abstract}
Multiple-input multiple-output (MIMO) is an enabling technology to meet the growing demand for faster and more reliable communications in wireless networks with a large number of terminals, but it can also be applied for position estimation of a terminal exploiting multipath propagation from multiple antennas. In this paper, we investigate new convolutional neural network (CNN) structures for exploiting MIMO-based channel state information (CSI) to improve indoor positioning. We evaluate and compare the performance of three variants of the proposed CNN structure to five NN structures proposed in the scientific literature using the same sets of training-evaluation data. The results demonstrate that the proposed residual convolutional NN structure improves the accuracy of position estimation and keeps the total number of weights lower than the published NN structures. The proposed CNN structure yields from 2\,cm to 10\,cm better position accuracy than known NN structures used as a reference.
\end{abstract}


%
\IEEEpeerreviewmaketitle

\section{Introduction}
\label{sec:intro}



Multiple-input multiple-output (MIMO) is an enabling technology to meet the growing demand for faster and more reliable wireless communications in wireless networks with a vast number of wireless terminals. The idea is to introduce space diversity through an increased number of antennas at the transmitter and/or receiver side, thus improving the wireless link reliability or increasing radio link capacity by exploiting multipath propagation.

The MIMO approach is already part of WiFi devices based on IEEE~802.11n standard. It is also extensively used in the third-generation (3G) and the fourth generation (4G) mobile broadband networks. Furthermore, the fifth-generation (5G) and future sixth-generation (6G) broadband networks are putting even more emphasis on multi-antenna technologies extending the concept to massive MIMO.

In addition to improving the reliability and capacity of wireless links, the multiple antennas can be exploited for indoor or outdoor positioning. Accurate indoor positioning (and position estimation in general) is a highly desirable feature of future wireless networks~\cite{lemic2016localization, wen2019survey5g}. It is a key enabler for a wide range of applications, including navigation, smart factories and cities, surveillance, security, IoT, sensor networks, and future reconfigurable intelligent surfaces (RIS). Additionally, indoor positioning can be leveraged for improved beamforming and channel estimation in wireless communications.

The positioning techniques can be classified into five classes; namely, i) proximity-based, ii) angle-based, iii) range-based, iv) fingerprinting-based and v) device-free localization \cite{savic2015fingerprinting, chin2020intelligent, Bregar2020}. The \textit{proximity-based} approaches rely on the information about which objects are detected in the observer vicinity. The \textit{angle-based} approaches rely on the angle-of-arrival (AoA) of the received signal as obtained by the multi-antenna system. The \textit{range-based} approaches rely on either time-of-arrival (ToA) or time-difference-of-arrival (TDoA) of the received signal or on received signal strength (RSS). Since the ToA approach requires large bandwidth (\eg 20\,MHz correspond to 15\,m accuracy), it is not always feasible. Next, the \textit{fingerprinting-based} approaches rely on the access to accurate channel state information (CSI), where the radio frequency (RF) fingerprints consist of signal measurements obtained at known positions within the deployment area. Fingerprint positioning can work with a single base station (BS), if CSI includes spatial channel information obtained from several antennas. Since the RF fingerprints typically form high-dimensional datasets, the fingerprinting-based localization is a good candidate for utilizing machine learning (ML) methods. Finally, in the \textit{device-free} localization, the system detects and tracks any entity based on an entity's impact on CSI.

The fingerprinting-based and the device-free localization classes rely on the CSI datasets that can be obtained by extensive measurement campaigns, which prove to be time consuming and expensive. Alternatively, the analytical and simulation-based approaches (\eg radio ray-traycing~\cite{novak2016discrete}) can supplement measurements to some extent. The simulated CSI accuracy, however, depends on the span of considered phenomena (see~\cite[p.~22]{wen2019survey5g}) and knowledge about a particular radio environment.


In this paper, motivated by CTW~2019 challenge~\cite{ctw2019flyer}, we investigate new machine learning approaches for improving the indoor positioning using CSI obtained by a single massive MIMO antenna. In particular, we propose a new convolutional neural network (CNN) structure with three variants in its internal layers, we compare its performance to some of the existing neural network (NN) structures on the same CSI dataset, and discuss possible further improvements.

The main contributions of this paper are the following:
\begin{itemize}
    \item the design of a new deep CNN structure able to accurately estimate the position of a transmitter in a room,
    \item the re-implementation of the NN structures published in scientific literature and comparison of their performance to the newl proposed CNN structure and
    \item the evaluation procedure with four different distributions of training/evaluation samples based on publicly available CSI dataset from the CTW~2019 challenge~\cite{ctw2019flyer}.
\end{itemize}

The rest of the paper is structured as follows. In Section~\ref{sec:setup}, we briefly analyze some representative studies on deep learning application for position estimation and provide the problem definition. Section~\ref{sec:dataset} describes the publicly available CTW~2019 challenge dataset, and Section~\ref{sec:model} outlines the new proposed CNN structure with its variants. Performance evaluation and comparison to five representative NNs is done in Section~\ref{sec:evaluation}. Discussion on lessons learned and conclusions are provided in Section~\ref{sec:conclusion}.

\section{Related Work and Problem Definition}
\label{sec:setup}

The use of ML methods for indoor positioning gained much attention from the research community in recent years, so we selected just a few studies that are relevant and representative to our approach. Savic and Larsson in~\cite{savic2015fingerprinting} focus on existing methods for position estimation using classical ML approaches, where they briefly present k-Nearest Neighbour, Support Vector Machine and Gaussian Process Regression as suitable candidates. Recent fingerprint-based positioning studies \cite{arnold2019localization,arnold2019sounding,bast2020positioning} consider larger antenna array and high number of considered subcarriers. A high number of antennas and subcarriers inevitably produce rich fingerprint samples and thus large final dataset size. Huge datasets make traditional ML approaches difficult to utilize, openning an opportunity to more recent ML approaches such as deep NNs that utilize either batch processing or stream processing on large datasets.

In the literature, we found several NN architectures proposed for more accurate position estimation. In~\cite{arnold2019localization, chin2020intelligent}, authors experiment with fully (\ie~densely) connected NNs (FCNN). However, feature extraction process using convolutional NNs (CNNs) proved far more effective from the perspective of the performance and the number of weights compared to FCNN~\cite{vieira2017fingerprinting, arnold2019practical, arnold2019sounding, widmaier2019towards, bast2020positioning, chin2020intelligent}. The top-performing NNs in related work use similar sequence of layers: a convolutional layer, an activation layer and an average pooling layer.

The main application of convolution layers is in image recognition domain. Images typically have sharp edges between surfaces, and from the gradient perspective they are much more ``dynamic'' compared to CSI, where changes are much slower as shown in Figure~\ref{fig:sample:csi} for the CTW2019 dataset. In the NN design, image-related tasks typically utilise convolutional layers with kernel shape $(n,n)$, while a significant part of related work for position estimation utilises kernel shape $(1,n)$.

Concerning position estimation accuracy, \cite{arnold2019localization, arnold2019sounding, bast2020positioning} show that an increasing number of antennas and therefore an amount of CSI improves the overall accuracy of position estimation. Furthermore, \cite{bast2020positioning} suggests generating additional features derived from raw CSI, in particular ``time series'' from inverse FFT and representation with polar coordinates, but in our study we did not obtain any performance gains when using additional features. Also, in \cite{bast2020positioning} authors show in their experiment that linear and rectangular antenna arrays perform slightly better in terms of accuracy than distributed antenna arrays around the testing area. However, their experiment was limited to a single plane, where height was unchanged, and event recording was done solely in front of the antenna arrays.

Several metrics that serve either as loss metrics at the training phase or as evaluation metrics are applied for performance evaluation. Most often used metrics in related work are mean distance error (MDE)~(\ref{eq:mde}), normalized mean distance error (NMDE)~(\ref{eq:nmde}) and root-mean-squared-error (RMSE)~(\ref{eq:rmse}). MDE gives the Euclidean distance error between the ground truth position $p$ and the estimated position $\hat{p}$. In ~(\ref{eq:mde}), $\norm{\cdot}_2$ thus stands for the Euclidean norm. NMDE is used because samples farther away from the antenna array are hard to estimate accurately. Thus, by normalizing, farther away samples receive less penalty for the error.


\begin{equation}\label{eq:mde}
  \textrm{MDE} = \mathbb{E}\left[ \norm{p - \hat{p}}_2 \right],
\end{equation}

\begin{equation}\label{eq:nmde}
  \textrm{NMDE} = \mathbb{E}\left[\frac{\norm{p - \hat{p}}_2}{\norm{p}_2}\right].
\end{equation}

\begin{equation}\label{eq:rmse}
  \textrm{RMSE} = \sqrt{ \mathbb{E}\left[ \norm{p - \hat{p}}_2^2 \right]}.
\end{equation}

In our study, the task of the applied ML approach was to predict the transmitter's location ($p = p(x,y,z)$) from the publicly available offline dataset~\cite{ctw2019dataset} containing CSI measurements at different positions. The proposed CNN structures were trained, to gain experience, and evaluated, to estimate the performance, on different non-overlapping subsets of the CSI dataset.

\section{Dataset Description}
\label{sec:dataset}

\begin{figure}[htbp]
  \centering
  \includegraphics[width=0.9\linewidth]{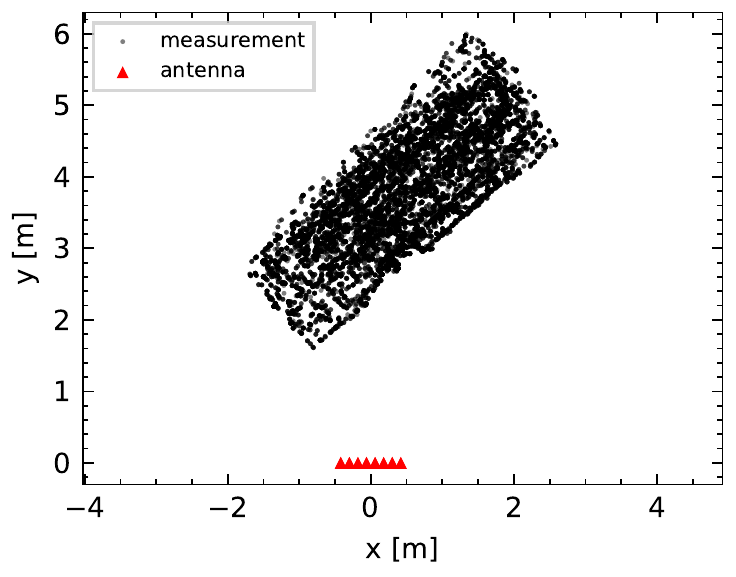}
  \caption{Top-down view on 17.486 samples and antenna orientation.}
  \label{fig:ctw2019:samples}
\end{figure}

As explained in Section~\ref{sec:setup}, for training of the newly proposed CNN and for the performance evaluation presented in Section~\ref{sec:evaluation} we used the openly available dataset from CTW~2019 challenge~\cite{ctw2019dataset}. The dataset was acquired by a massive MIMO channel sounder~\cite{arnold2019sounding} in a setup visually depicted in~\cite{ctw2019flyer}. The CSI was measured between a moving transmitter and $8\times2$ antenna array. The transmitter implemented on SDR was placed on a vacuum-cleaner robot. The robot drove in a random path on approximately 4\,m~$\times~2$\,m size table. The transmitted signal consisted of OFDM pilots with a bandwidth of 20\,MHz and 1024 subcarriers at the central frequency 1.25\,GHz. 100 sub-carriers were used as guard bands, 50 on each side of the frequency band. The measurement setup is summarized in Table~\ref{tab:ctw2019specs}, while Figure~\ref{fig:ctw2019:samples} depicts the top-down view on the the antenna orientation and positions of 17.486 CSI samples available in the dataset.

\begin{table}[htbp]
\ra{1.3}
\caption{Dataset Summary}
\label{tab:ctw2019specs}
\centering
\begin{tabular}{@{}ll@{}}
  \toprule
  Property & Value \\
  \midrule
  No. antennas & 16 \\
  Central frequency ($f_{c}$) & 1.25\,GHz \\
  Bandwidth & 20\,MHz \\
  No. subcarriers & 924 useful; $\approx$\,20\,kHz in between \\
  CSI data shape & 16 (ant.) $\times$ 924 (subc.) $\times$ 2 (Re/Im) \\
  SNR information & 16 (ant.) $\times$ 1 SNR value given in dB \\
  No. all samples & 17\,486 samples \\
  \bottomrule
\end{tabular}
\end{table}

\begin{figure}[htbp]
  \centering
  \subfloat[Real part of channel response\label{fig:sample:real}]{\includegraphics[width=0.9\linewidth]{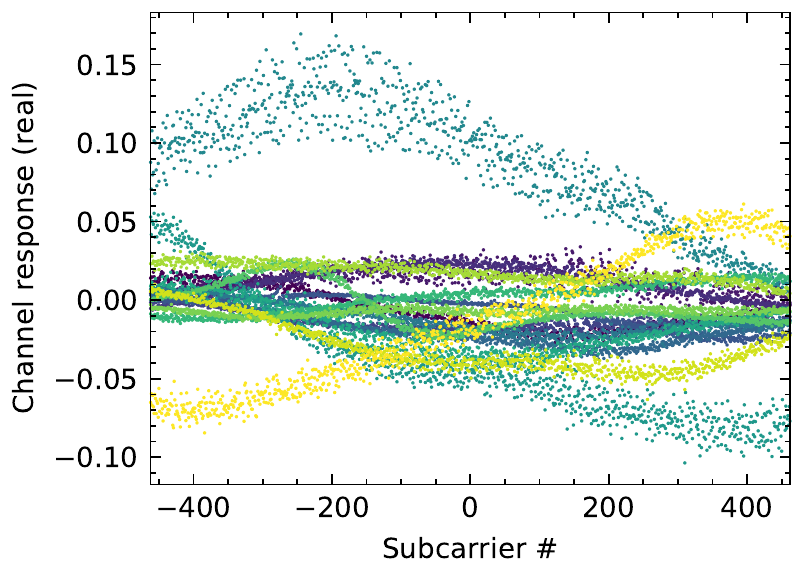}}\\
  \subfloat[Imaginary part channel response\label{fig:sample:imag}]{\includegraphics[width=0.9\linewidth]{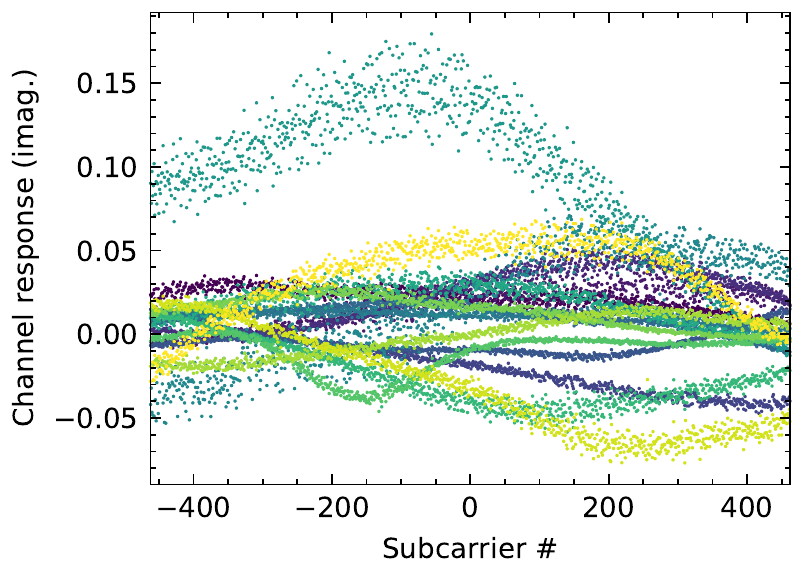}}
  \caption{Channel State Information of sample \#130}
  \label{fig:sample:csi}
\end{figure}

As an example, the CSI sample \#130 is presented in Figure~\ref{fig:sample:csi}, namely, real part in  Figure~\ref{fig:sample:real} and imaginary part in Figure~\ref{fig:sample:imag}. Both figures present sub-carriers ordered linearly (by their frequency) from left to right, where the value on x-axis present padding (approx. 20\,kHz) from the central frequency (1.25\,GHz). Each of the seemingly continuous curves (despite slight value deviations) distincted by colour presents individual channel responses obtained by one of the 16 antennas.

Since we are dealing with a limited amount of data, we divided the dataset to non-overlapping training and evaluation subsets. Moreover, to investigate the influence of training data selection, we generated four different training/evaluation sets from the same original dataset using (i) a random dataset split with overlapping training and evaluation areas, (ii) a long-edge dataset split with narrow evaluation area, (iii) a short-edge dataset split with wide evaluation area, and (iv) a cut-out dataset split with evaluation area within the training area. These four approaches are graphically depicted in Figures~\ref{fig:validation}a, b, c and d, respectively. All approaches split the dataset to training and evaluation subsets approximately at a 9:1 ratio.

\section{Proposed Neural Network Structures}
\label{sec:model}

In this paper, we propose three CNN structures with multiple layers designed for position estimation based on CSI samples. Unless otherwise stated, after each layer except the last one, we utilize a rectified linear unit (ReLU) activation. The first structure denoted as CNN4 uses four sequential convolutional layers with the kernel shape $(1,7)$ and stride $(1,3)$, which are followed by a dense layer with 1000 units. With each convolutional layer, the number of filters (depth) is increased by 50\%.

The second structure denoted as CNN4R is similar to CNN4, but instead of four convolutional layers, it utilizes four residual network (ResNet) blocks inspired by ResNet-18~\cite{he2016resnet}. CNN4R's ResNet blocks use kernel size $(1,7)$, where each block reduces width with stride $(1,3)$ and internal, pattern with identity connection repeats three times.

The third structure denoted as CNN4S is based on CNN4R. However, instead of the first ResNet block it utilizes a single convolutional layer (\ie stem) with kernel $(1,7)$ and stride $(1,2)$, followed by an average pooling layer with a pool size of $(1,4)$ and stride $(1,2)$, which functions as a rolling average.

\section{Performance Evaluation}
\label{sec:evaluation}

\begin{table*}
\ra{1.3}
\caption{Performance evaluation on CTW-2019 dataset}
\label{tab:evaluation}
\centering
\begin{tabularx}{\linewidth}{@{}Xr|rrr|rrr|rrr|rrr@{}}
    \toprule
    \multirow{2}{*}[-0.5em]{Approach}
    & \multirow{2}{*}[0.5em]{Weights}
    & \multicolumn{3}{c|}{Random}
    & \multicolumn{3}{c|}{Narrow}
    & \multicolumn{3}{c|}{Wide}
    & \multicolumn{3}{c}{Within}
    \\
    
    \cmidrule(lr){3-5}
    \cmidrule(lr){6-8}
    \cmidrule(lr){9-11}
    \cmidrule(l){12-14}
    
    {} & $[10^6]$
    & RMSE & MDE & NMDE
    & RMSE & MDE & NMDE
    & RMSE & MDE & NMDE
    & RMSE & MDE & NMDE
    \\
    \midrule
    
    
    Dummy (linear), FCNN
    & <0.1
    & 0.724 & 1.122 & 25.1
    & 1.055 & 1.809 & 51.4
    & 0.878 & 1.428 & 28.2
    & 0.441 & 0.721 & 15.1
    \\
    
    Arnold~\etal~\cite{arnold2019localization}, FCNN
    & 32.3
    & 0.570 & 0.853 & 19.4
    & 1.001 & 1.594 & 45.0
    & 0.733 & 1.145 & 23.3
    & 0.381 & 0.584 & 12.3
    \\
    
    Arnold~\etal~\cite{arnold2019sounding}, CNN
    & 7.6
    & 0.315 & 0.445 & 10.0
    & 0.857 & 1.330 & 37.7
    & 0.605 & 0.923 & 18.6
    & 0.454 & 0.702 & 14.8
    \\
    
    Bast~\etal~\cite{bast2020positioning}, CNN
    & 0.4
    & 0.722 & 1.120 & 25.1
    & 1.110 & 1.907 & 54.2
    & 0.828 & 1.331 & 26.5
    & 0.377 & 0.611 & 12.7
    \\
    
    Chin~\etal~\cite{chin2020intelligent} FCNN
    & 123.6
    & 0.563 & 0.838 & 19.0
    & 1.007 & 1.611 & 45.4
    & 0.726 & 1.133 & 23.0
    & 0.365 & 0.574 & 12.0
    \\
    
    Chin~\etal~\cite{chin2020intelligent} CNN
    & 13.7
    & 0.100 & 0.093 & 2.1
    & 0.854 & 1.326 & 37.8
    & 0.530 & 0.808 & 16.3
    & 0.381 & 0.620 & 13.0
    \\
    
    \midrule
    
    CNN4 
    & 5.3
    & 0.122 & 0.149 & 3.4 
    & 0.819 & 1.286 & 36.6 
    & 0.514 & 0.787 & 15.9 
    & 0.365 & 0.552 & 11.6 
    \\
    
    CNN4R 
    & 10.8
    & 0.113 & 0.127 & 2.8 
    & 0.776 & 1.227 & 34.7 
    & 0.539 & 0.835 & 16.8 
    & 0.351 & 0.521 & 11.0 
    \\

    CNN4S 
    & 16.3
    & 0.108 & 0.120 & 2.7 
    & 0.821 & 1.285 & 36.5 
    & 0.528 & 0.804 & 16.2 
    & 0.351 & 0.524 & 11.1 
    \\
\bottomrule
\end{tabularx}
\end{table*}

In order to benchmark the proposed CNN structures, we also implemented some representative NNs for position estimation from the related work~\cite{arnold2019localization, arnold2019sounding, bast2020positioning, chin2020intelligent}. The evaluation results are presented in Table~\ref{tab:evaluation}, where RMSE and MDE are expressed in meters, and NMDE is given in per-cents. In general, the CNN structures exhibit better performance compared to FCNN. An exception is~\cite{sobehy2019ndr}, where authors had put significant effort into data pre-processing, but unfortunately we were unable to replicate the authors' results.

\begin{figure*}[htbp]
  \centering
  \subfloat[Random dataset split\label{fig:validation:random}]{\includegraphics[width=0.3\linewidth]{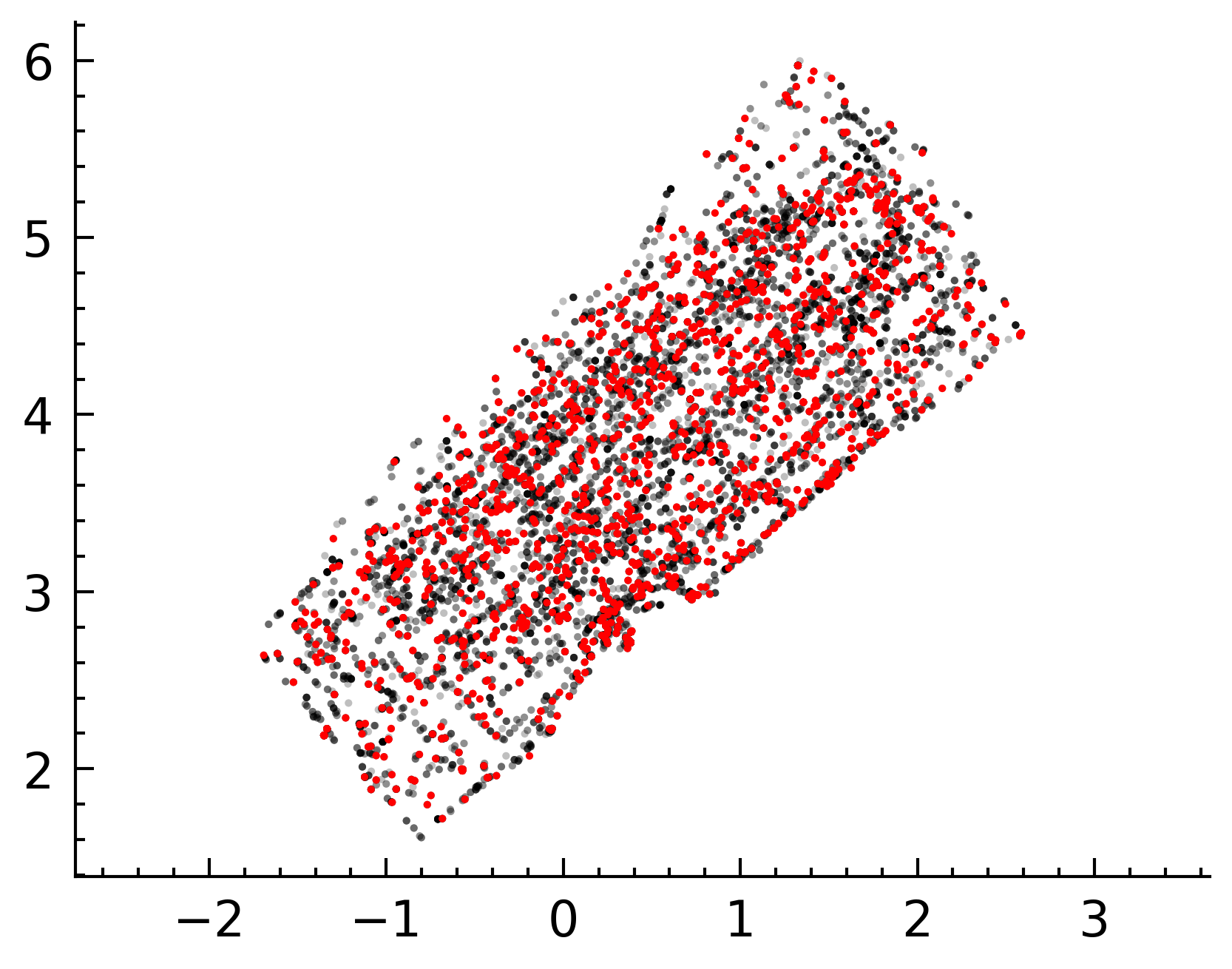}}\hfil%
  \subfloat[Narrow evaluation area\label{fig:validation:narrow}]{\includegraphics[width=0.3\linewidth]{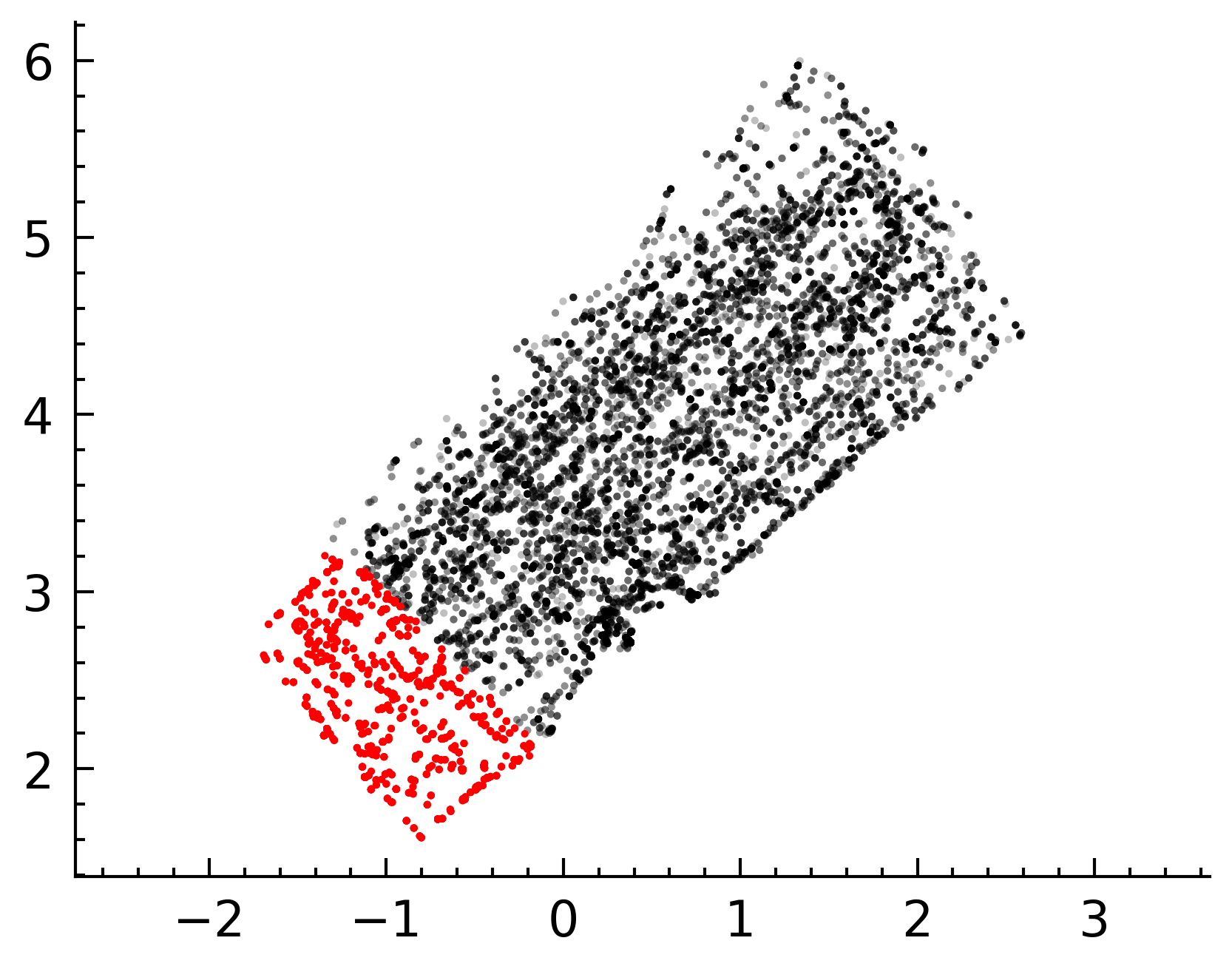}}

  \subfloat[Wide evaluation area\label{fig:validation:wide}]{\includegraphics[width=0.3\linewidth]{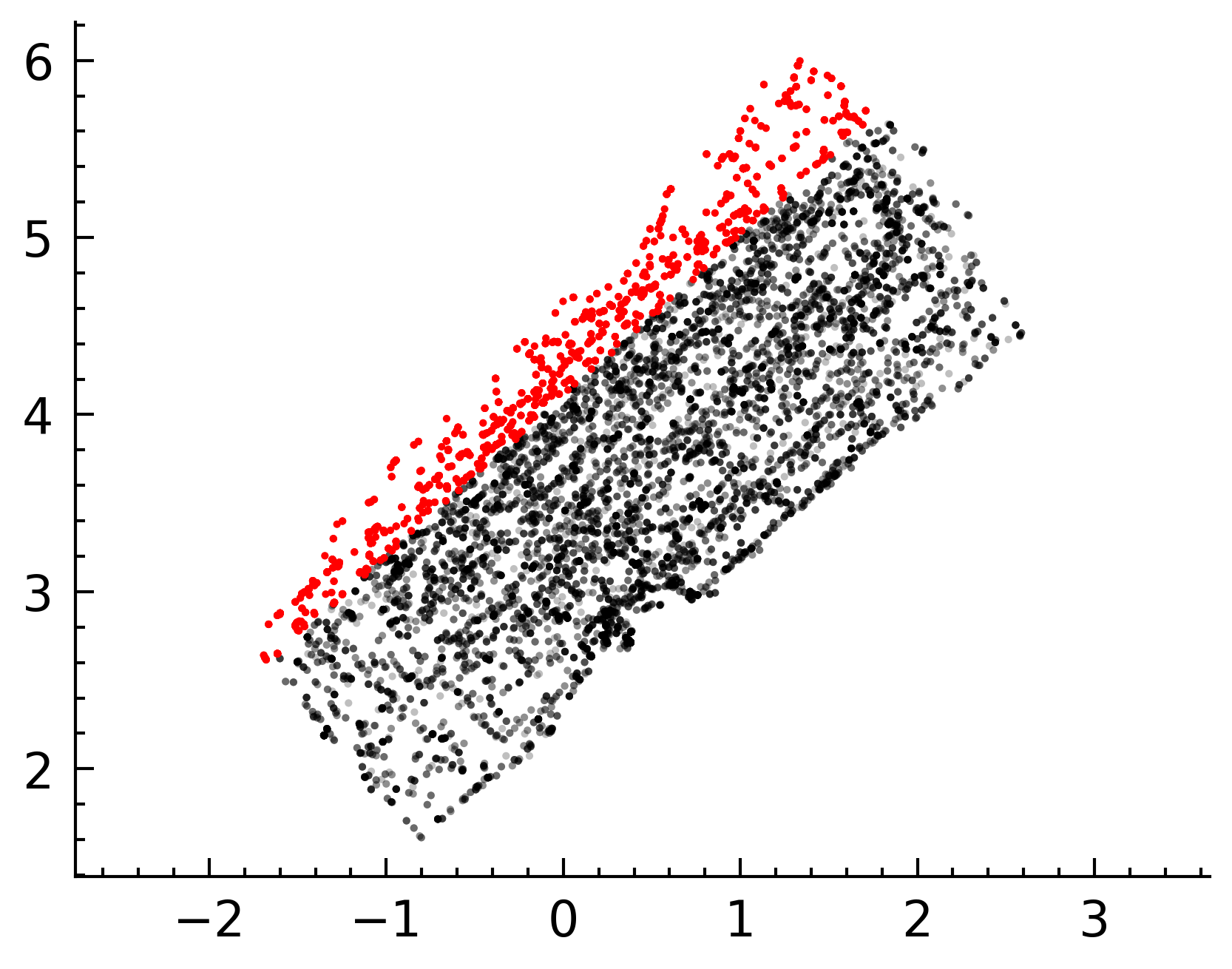}}\hfil%
  \subfloat[Cut-out evaluation area\label{fig:validation:within}]{\includegraphics[width=0.3\linewidth]{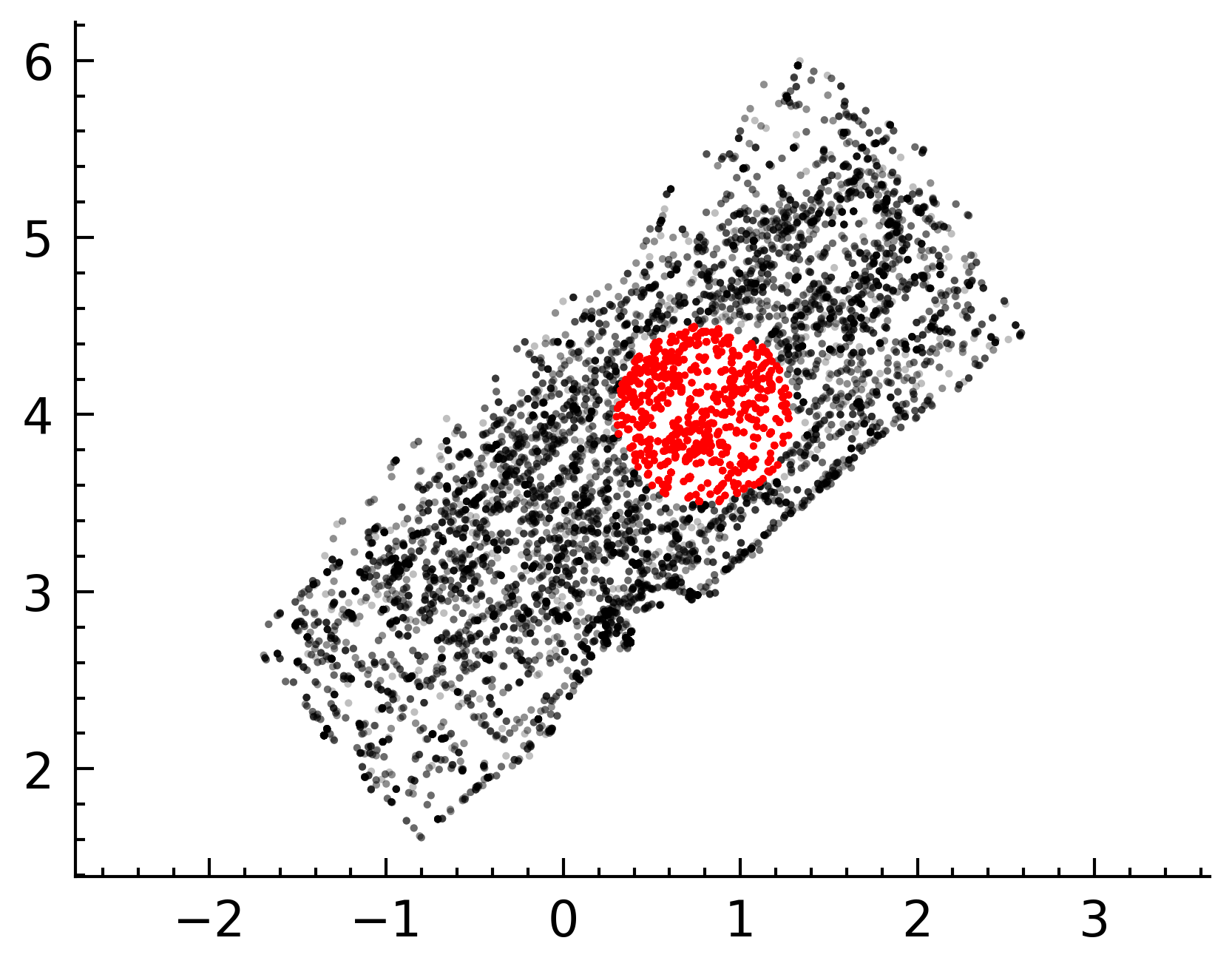}}
  \caption{Four types of training/evaluation data subsets, where training data is marked black and testing data is marked red.}
  \label{fig:validation}
\end{figure*}



Each experiment ran for 250 epochs at most, where we considered early stopping (after 21 epochs without improvement) and exponential decrease of learning rate (\ie $\textrm{rate}\leftarrow\textrm{rate}/10$) after 10 epochs without improvement. The batch size was 32, and MDE was used as a training loss metric. Weights were updated using stochastic gradient descent with a learning rate of $10^{-3}$ and 0.9 momentum.


\begin{figure*}[htbp]
    \centering
    \subfloat[Bast~\etal~\cite{bast2020positioning}, CDF\label{fig:cdf:bast2020}]{\includegraphics[width=.33\linewidth]{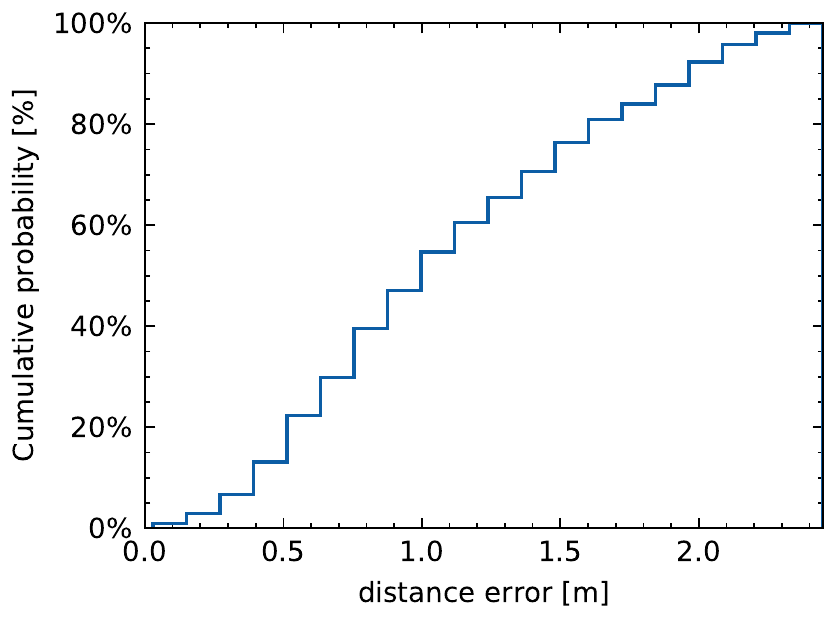}}
    \subfloat[Bast~\etal~\cite{bast2020positioning}, XY histogram\label{fig:hist:bast2020}]{\includegraphics[width=.31\linewidth]{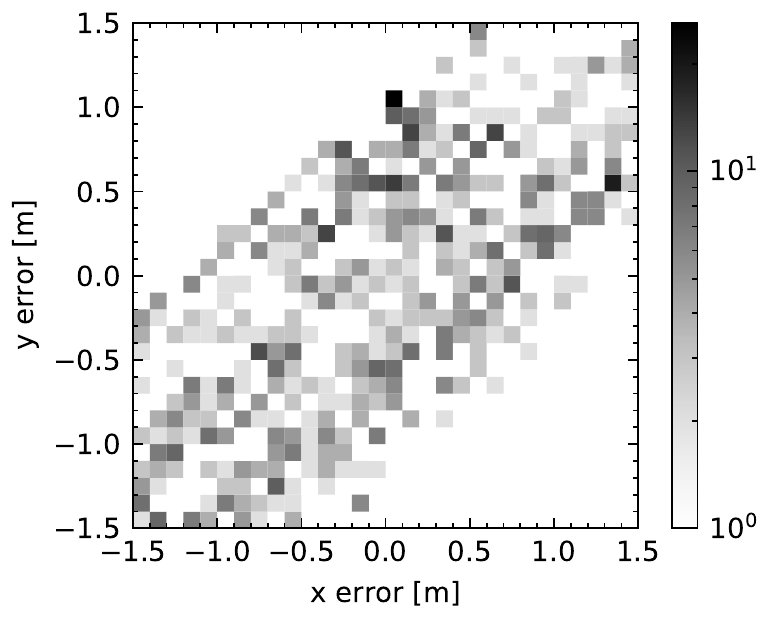}}
    \subfloat[Bast~\etal~\cite{bast2020positioning}, QuiverPlot\label{fig:quiver:bast2020}]{\includegraphics[width=.31\linewidth]{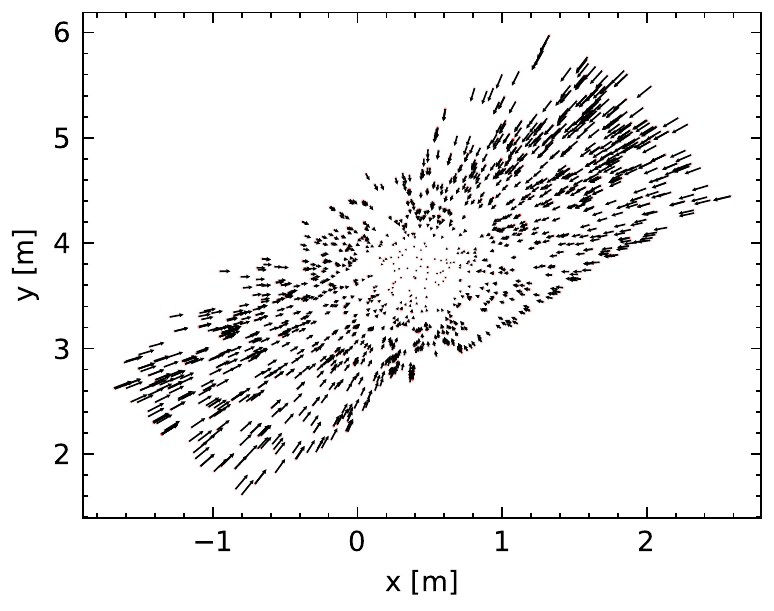}}\\ 
    \subfloat[CNN4, CDF\label{fig:cdf:variant1}]{\includegraphics[width=.33\linewidth]{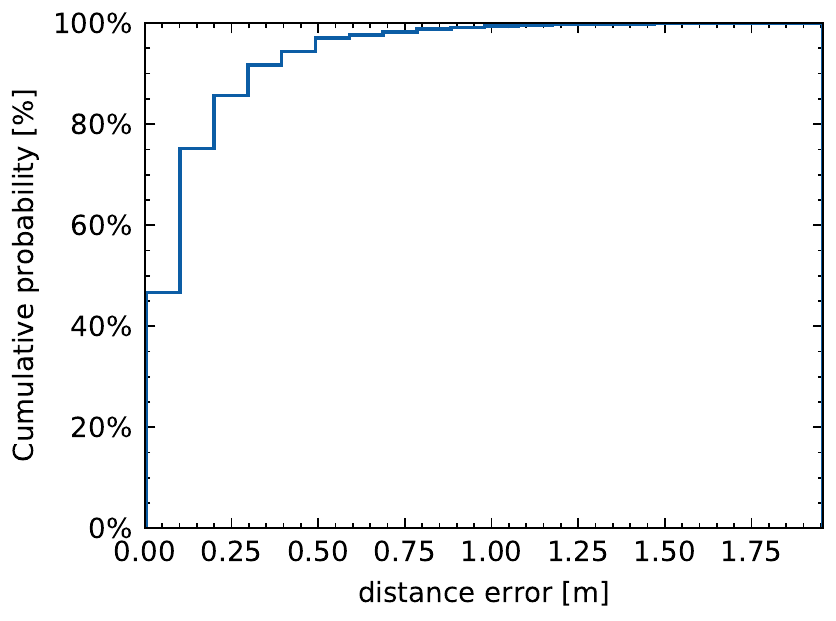}}
    \subfloat[CNN4, XY histogram\label{fig:hist:variant1}]{\includegraphics[width=.31\linewidth]{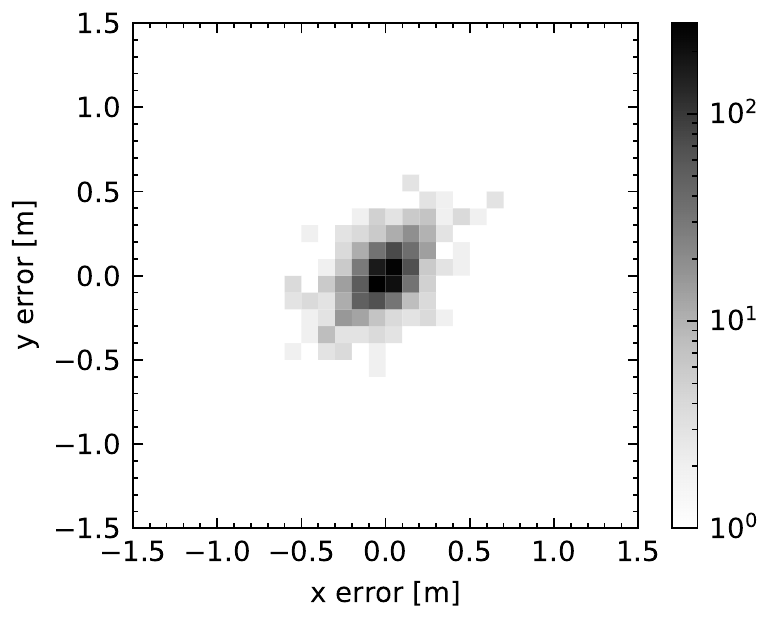}}
    \subfloat[CNN4, QuiverPlot\label{fig:quiver:variant1}]{\includegraphics[width=.31\linewidth]{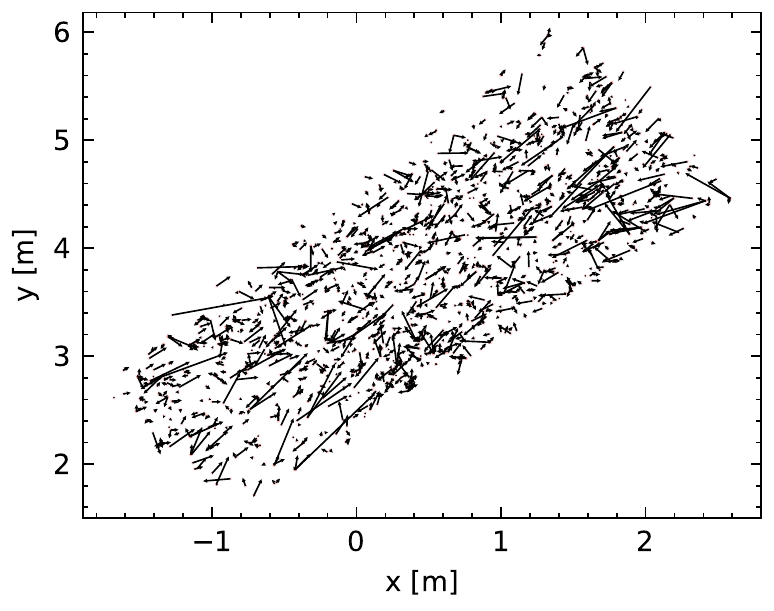}}\\
    \subfloat[CNN4R, CDF\label{fig:cdf:variant2}]{\includegraphics[width=.33\linewidth]{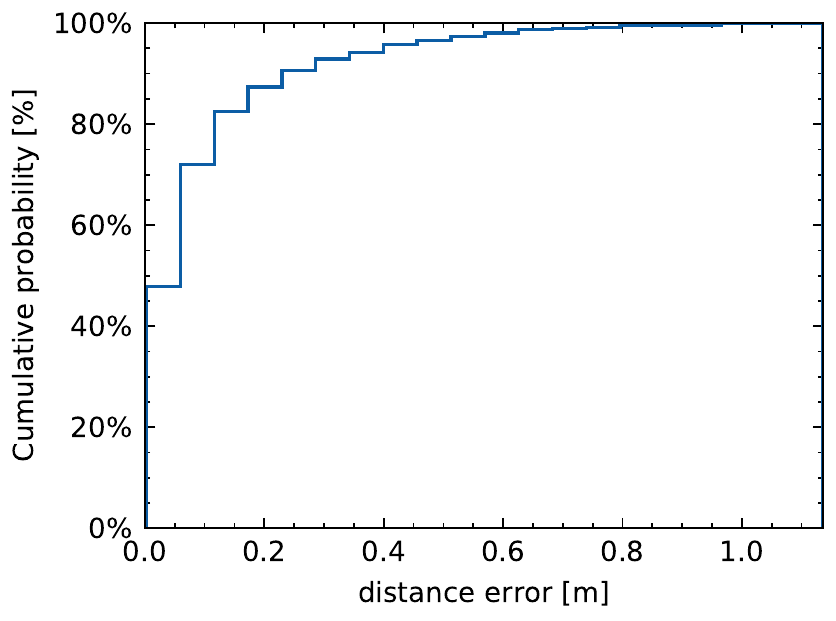}}
    \subfloat[CNN4R, XY histogram\label{fig:hist:variant2}]{\includegraphics[width=.31\linewidth]{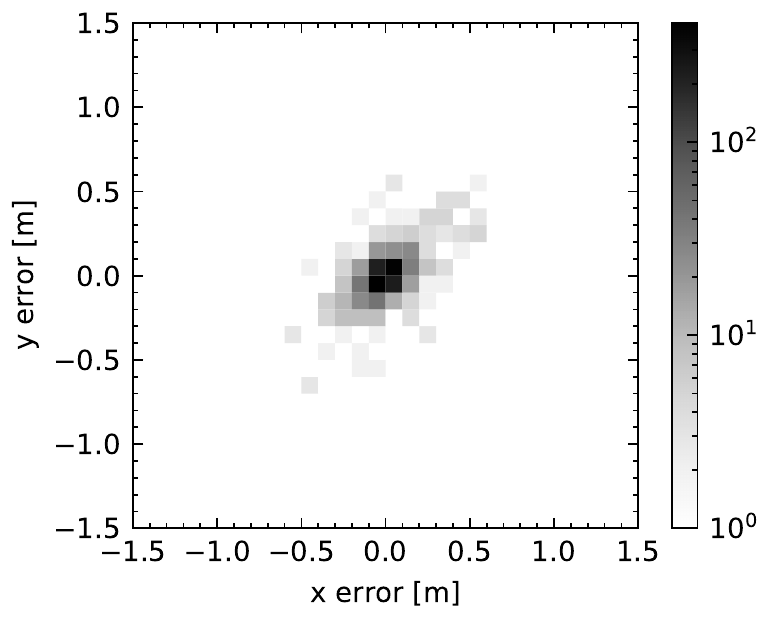}}
    \subfloat[CNN4R, QuiverPlot\label{fig:quiver:variant2}]{\includegraphics[width=.31\linewidth]{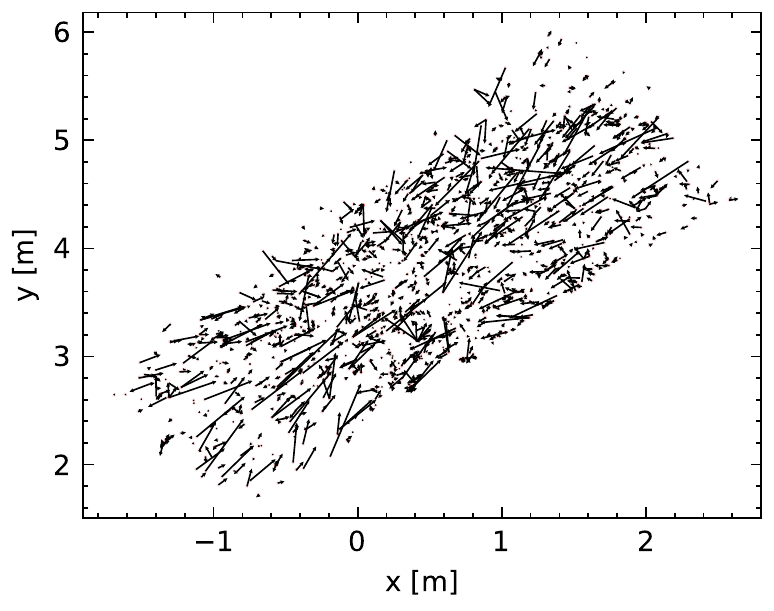}}\\
    \subfloat[CNN4S, CDF\label{fig:cdf:variant3}]{\includegraphics[width=.33\linewidth]{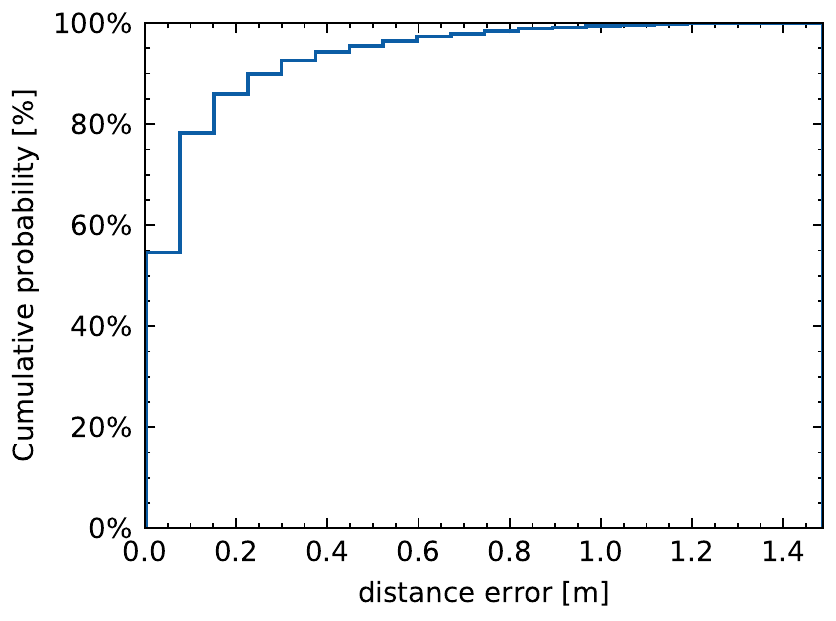}}
    \subfloat[CNN4S, XY histogram\label{fig:hist:variant3}]{\includegraphics[width=.31\linewidth]{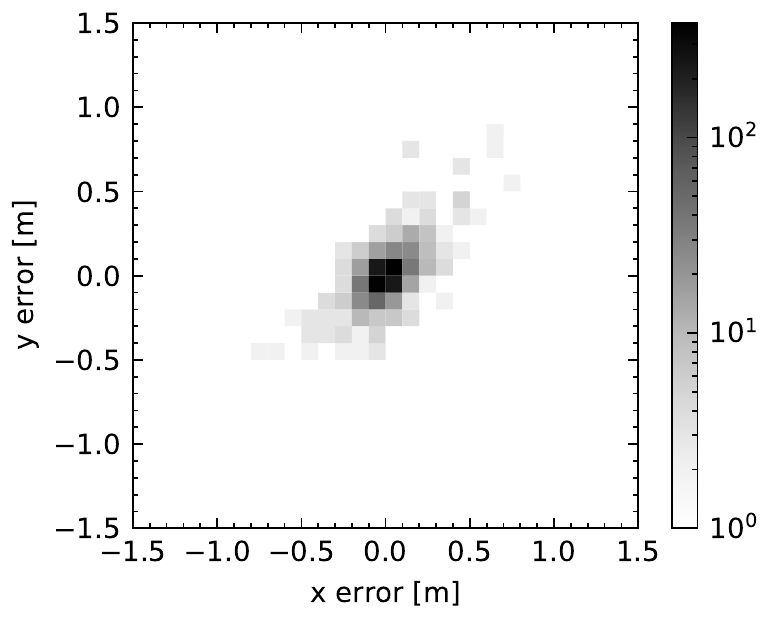}}
    \subfloat[CNN4S, QuiverPlot\label{fig:quiver:variant3}]{\includegraphics[width=.31\linewidth]{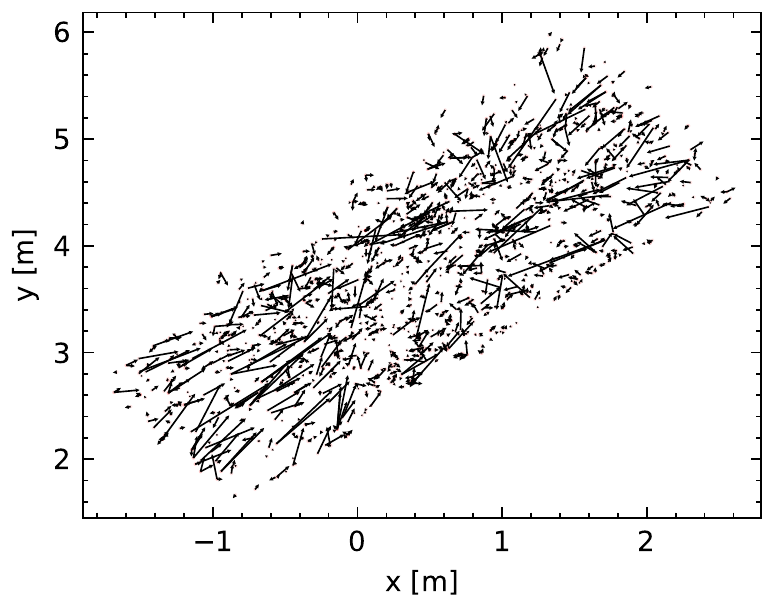}}
    
    \caption{Error distribution for proposed CNN structures and a reference Bast~\etal~\cite{bast2020positioning} structure presented in terms of CDFs of estimation error, histogram of error along X and Y axis, and quiver plot with arrows pointing from ground truth values towards error offset. Note that quivers were automatically scaled.}
    \label{fig:error-distribution:random}
\end{figure*}

The results in Table~\ref{tab:evaluation} show that the fingerprinting approach works best when the training set and evaluation set overlap (i.e. in random dataset split shown in Fig.\ref{fig:validation:random}). Because training and evaluation samples are close to each other, the results do not include error due to unbalanced training. The performance difference between NN structures is also the most significant for this dataset split. We see that difference between the best and the worst performing model is almost one meter or 23 per cent for NMDE. For this scenario, the best performing model is CNN~\cite{chin2020intelligent}, but our proposed CNN structures show comparable performance. However, in the case of dataset with narrow evaluation area (Fig.~\ref{fig:validation:narrow}), our model performs slightly better than CNN~\cite{chin2020intelligent}. The difference in MDE is up to 6\,cm. Even when the models are evaluated by dataset with wide validation area (Fig.~\ref{fig:validation:wide}), our model performs slightly better than CNN~\cite{chin2020intelligent}, but the difference is only up to 2\,cm. Additionally, we see that our model performs slightly better when evaluated by dataset with cutout area (Fig.~\ref{fig:validation:within}). The estimated distance error difference with respect to CNN~\cite{chin2020intelligent} is up to 3\,cm.

In Figure~\ref{fig:error-distribution:random}, we graphically present the error distribution of Bast~\etal~\cite{bast2020positioning} approach and three newly proposed structures for random train/evaluate scenario. We present position estimation accuracy using three types of figures. The figures in first column present cumulative distribution function (CDF), the second column contains figures with discrete error density over X and Y axis, and in the third column, we depicted quiver plot, where quivers/arrows points from ground truth toward the estimated position in relative scale. Figures~\ref{fig:cdf:bast2020} and \ref{fig:hist:bast2020} show that NN structure proposed by Bast~\etal~\cite{bast2020positioning} has error distribution far more spread compared to our proposed structures. Figure~\ref{fig:quiver:bast2020} reveals that Bast~\etal~\cite{bast2020positioning} structure has decent accuracy at the centre of the dataset other points have a bias toward the centre. CNN4R structure shows the highest accuracy, which is also aligned with Table~\ref{tab:evaluation}. The MDE values are below 1.2\,m (Fig.~\ref{fig:cdf:variant2}). The same can be observed by looking at Figure~\ref{fig:hist:variant2}, where the histogram is narrower compared to the other two proposed models. In addition, analysis of the quiver plots in the third column in Figure~\ref{fig:error-distribution:random} shows that there are no inconsistent areas present.

\section{Conclusions and Future Work}
\label{sec:conclusion}




In this paper we proposed three new CNN structures referred to as CNN4, CNN4R and CNN4S, designed for improving indoor positioning based on CSI obtained from a single massive MIMO antenna. The performance evaluation of the CNN structures shows that they are within the top ranked or the best performers in all given scenarios for different training/evaluation datasets derived from the publicly available CSI dataset from the CTW 2019 challenge~\cite{ctw2019flyer}. Even though the performance difference is minimal, or in some cases even in the noise range, we achieve similar performance to NNs~\cite{arnold2019sounding,chin2020intelligent} with a significantly lower number of trainable weights. Since we utilize strides instead of pooling operation, the computational cost is also lower than that of the comparable models.

The proposed CNN structures as well as those from the related work are not optimal. For our CNNs, we focused solely on pursuing the highest position estimation accuracy. Thus, in the process, we ignored several vital aspects of NNs that may be worth of further investigation, such as dead neurons, extreme weight values and fading/exploding gradient. While dead neurons could be prevented by using different activation functions, they can pose an opportunity for pruning NN, which would reduce the number of weights, making it sparse and consequently decreasing the neural network's size. The extreme weight values can be tackled using regularisation, but it would significantly increase the training time and the number of tunable parameters. To adequately address the fading/exploding gradient, besides using the residual connection, we see great potential in the recently introduced self-normalised NNs. However, their full potential has yet to be explored.

\section*{Acknowledgments}
The Slovenian Research Agency supported this work under grants P2-0016 and J2-2507. The authors would like to thank Maximilian Arnold from the University of Stuttgart, one of the CTW dataset creators, for kindly responding to our questions during the initial exploration of the dataset.

\ifCLASSOPTIONcaptionsoff
  \newpage
\fi

\bibliographystyle{IEEEtran}
\bibliography{research}

\end{document}